# XANES determination of chromium oxidation states in glasses: comparison with optical absorption spectroscopy

Olivier Villain, Georges Calas, Laurence Galoisy, Laurent Cormier

Institut de Minéralogie et de Physique des Milieux Condensés (IMPMC), 75015 PARIS

Jean-Louis Hazemann

Laboratoire de Cristallographie, 38000 GRENOBLE

**Abstract**

The oxidation state of chromium in glasses melted in an air atmosphere with and without refining agents was investigated by Cr K-edge X-ray Absorption Near-Edge Structure (XANES) and optical absorption spectroscopy. A good agreement in the relative proportion of Cr(III) and Cr(VI) is obtained between both methods. We show that the chemical dependence of the absorption coefficient of Cr(III) is less important in XANES than in optical absorption spectroscopy. The comparison of glasses melted under different conditions provides an indirect assessment of the molar extinction coefficient of Cr(VI) in glasses.

# 1. Introduction

Chromium is largely used in glasses for a broad range of applications. The redox state of Cr provides glasses specific light transmission properties, which are used, e.g. in packaging technology. Cr(III) gives a green coloration in silicate glasses, whereas Cr(VI) imparts a yellow color [1] and provides interesting thermochromic properties [2]. Cr(III) is also an efficient nucleating agent in oxide glasses [3] and the redox state controls Cr solubility in glasses and melts [4]. Glass melting conditions are then adjusted to control Cr redox state, through an adequate choice of melting atmosphere or refining agents [5]. The enforcement of the Packaging Directive 62/94 has required the development of reliable reference methods for the quantification of chromium, owing to the environmental impact of chromate species. Chemical analyses is recommended for the determination of the concentration of Cr(VI) and total chromium ($Cr_{total}$)[6]. A disadvantage of chemical methods is that structural information and hence speciation are lost. In addition, during chemical dissolution of the glass, the $Cr(VI)/Cr_{total}$ ratio may be modified by the presence of other redox pairs. Various spectroscopic methods have been alternatively used for assessing Cr-oxidation state. Optical absorption spectroscopy is widely used for Cr content and Cr redox determinations [1], but may be limited by the overlap with the absorption bands from other transition elements. EPR has also been used to determine the Cr(III) concentration in glasses [7], but the Cr(III) signal may be hidden by the presence of other paramagnetic species. XPS has been used to measure the redox state of chromium at the glass surface [8]. Finally, XANES has been extensively used in the evaluation of the oxidation state of transition elements [9]. Cr-speciation is often determined by Cr K-edge XANES in natural and synthetic environmental materials ([10], [11]). Only a few studies report XANES-based Cr redox measurements in partially oxidized glasses. Borosilicate glasses of known $Cr(VI)/Cr_{total}$ ratio were used as references for XANES measurements of various compounds [12]. A qualitative detection of the presence of Cr(VI) by XANES was also performed in a soda-lime-silicate glass [13]. However, no quantitative comparison of the analysis of Cr redox states in glasses has been published so far between XANES, chemical methods and other spectroscopic techniques.

This study discusses the application of XANES spectroscopy to determine the oxidation state of Cr in glasses. Glasses of various compositions have been investigated, covering a wide range of Cr(III) and Cr(VI) ratio, and different $Cr_{total}$ concentrations. The $Cr(VI)/Cr_{total}$ ratios are in agreement with those obtained by optical absorption spectroscopy. The chemical dependence of the absorption coefficient of Cr(III) is less important in XANES than in optical absorption spectroscopy. Molar extinction coefficient values can be constrained by XANES results.

## 2. Experimental Procedure

The glasses were prepared by melting reagent grade oxides and sodium carbonate in a platinum crucible for 1h15 at 1500°C (1100°C for borate glass) in an air furnace (Table I). Refining agents like $As_2O_3$ and $SO_3$ ([5]) were used for SN, SCN and borosilicate glasses to limit or even avoid Cr(III) oxidation in Cr(VI). Glasses were cast and annealed at 625°C during 1h. Glass samples were checked for homogeneity and amorphous state by XRD. Their composition was obtained by ICP-AES (Table I) and the chemical homogeneity was verified by electron microprobe analysis. Room-temperature UV-visible-NIR transmission spectra were recorded using a double-beam Cary 5 spectrometer on polished glass slices. After correction for reflection, the absorption spectra were normalized to sample thickness.

Measured density and average molecular weights of the glasses were used to calculate Cr(III) and/or Cr(VI) concentration.

Fluorescence XANES spectra were recorded at room temperature on the FAME beamline (BM30B) of the European Synchrotron Radiation Facility (Grenoble, France). Cr K-edge fluorescence detection used a liquid nitrogen-cooled pseudo channel-cut Si(220) double-crystal monochromator and a 30-elements Ge detector. At the focus, the beam dimensions were 300 μm horizontal by 200 μm vertical. The energy resolution was 0.3eV [14]. Calibration was made with respect to the first inflection point in a Cr metal foil (5989.0 eV). ($KCr(SO_4)_2.12H_2O$) and $K_2Cr_2O_7$, mixed with cellulose and mounted on a Kapton tape, were used as references for Cr(III) and Cr(VI), respectively. XANES spectra were normalized using a second-degree polynomial function from 5900eV to 5950eV. Normalization was achieved by extrapolating into the edge region a fit to the data in the post-edge region from 6150 to 6400eV. The pre-edge features were extracted by subtracting an arctan function to fit the background [15].

## 3. Results and Discussion

A comparison between the spectra of SBN and SBNox glasses, melted under reducing and oxidizing conditions, respectively, shows the presence of the Cr(VI) band around 28000 cm$^{-1}$, in the oxidized glass (Fig. 1). The tail of this band is superimposed to the Cr(III) absorption bands and precludes accurate measurements in glasses, in which Cr(VI) is the major Cr oxidation state. The relative intensity of the Cr(VI) absorption band increases from SBNox to SCN, SN2ox and BN glasses. By contrast, SN, SBN and SBN2 glasses contain only Cr(III) (Fig. 1). The Cr(III) molar extinction coefficients ε has been taken from the absorption maximum of the crystal-field transition band (Table I). The differences observed between ε values may arise from variations in Cr(III)-site geometry or in the covalence of the Cr-O bond, a parameter which varies by about 10% with glass composition [17]. These variations preclude the use of a unique ε standard value for quantification of Cr-redox state in glasses. However, ε has a similar value in SN and other sodium silicate glasses [16, 18, 19], in which it ranges between 17 and 20 L mol$^{-1}$ cm$^{-1}$.

The background-subtracted, normalized Cr-K edge XANES spectra are dependent on the Cr-oxidation state (Fig. 2). A major characteristic of the XANES spectrum of Cr(VI) is a sharp pre-edge feature located approximately 15 eV lower in energy than the main absorption edge. The intensity of the Cr(VI) feature increases from SBN to SN2ox and BN glasses, which is consistent with optical absorption spectroscopy. In SBNox, the area of the pre-edge feature is 1.5 time higher than in SBN (Table I). This Cr(VI) component is prevailing in oxidized glasses, SN2ox and BN, hiding the coexisting Cr(III) pre-edge feature. However, Cr is not fully oxidized in the investigated glasses, as shown by the comparison with the Cr(VI) reference (inset of Fig. 2). The Cr(VI) pre-edge feature is not observed in the most reduced glasses such as SBN, indicating the absence of Cr(VI) in these samples. The Cr(III) pre-edge feature of SN glass has a normalized intensity of about 0.06 (Fig.3). It presents two components corresponding to the Cr(III) 3d-levels split by an octahedral crystal field (Villain et al, in prep). In the Cr(III) reference, the pre-edge presents the same shape, with the low-energy component being the most intense. However, the pre-edge intensity is twice higher for glasses than for the Cr(III) reference (Fig.3), due to stronger d-p mixing or site distortion. SBN and SBN2 glasses present a similar pre-edge feature, with a higher intensity (Fig.3). The specific shape and relative intensity of the Cr(III) pre-edge imply the use of glasses as Cr(III) references for the calculation of the redox state of Cr. In moderately oxidized glasses, SBNox and SCN, the feature at about 5994 eV is the most intense, which corresponds to the position of the Cr(VI) pre-edge.

The Cr(VI)/Cr$_{total}$ ratio has been determined using a linear combination of the pre-edge of the Cr(VI) reference and Cr(III) glass. Due to the chemical dependence of the pre-edge spectra, SBN and SN pre-edge features have been taken as references for SBNox and silicate glasses (i.e. SN2ox, SCN), respectively. The Cr(VI)/Cr$_{total}$ ratio determined by XANES is reported in Table 1. For SCN glass, the Cr(VI)/Cr$_{total}$ ratio obtained by XANES with SN as the Cr(III) reference is similar to the value obtained by optical absorption with $\varepsilon_{Cr(III)}$ equal to 18 L. mol$^{-1}$.cm$^{-1}$ (after [1]). In sodium silicate glasses, using the $\varepsilon_{Cr(III)}$ value determined in the reduced SN glass, the Cr(VI)/Cr$_{total}$ ratio is 30±7% in the SN2ox glass using optical absorption spectroscopy, a value consistent with XANES data. The values are in good agreement with XPS measurements on sodium silicate glasses of similar composition [8]. Higher values were obtained by chemical analyses and optical absorption, probably due to different redox state of the melt [20].

In borosilicate glasses, the redox values obtained for SBNox from XANES and optical absorption are consistent, using SBN as a Cr(III) reference for both methods (Table 1). The choice of the Cr(III) reference is crucial at low Cr(VI)/Cr$_{total}$ ratio, especially when using optical absorption spectroscopy. Relative to $\varepsilon_{Cr(III)}$ in SN, $\varepsilon_{Cr(III)}$ is higher by a factor of 1.7 and 1.2 in SBN and SBN2, respectively. By comparison, the area of XANES pre-edge peaks is similar in SBN and SBN2 and larger than in SN by a factor of 1.4. Consequently, XANES results are less dependant to the choice of the Cr(III) reference than optical absorption results.

In borate glasses, the Cr(VI)/Cr$_{total}$ ratio measured by XANES is 60±5 % in BN glass, irrespective of Cr(III) reference, due to the major contribution of the Cr(VI) feature (Table 1). This redox ratio is in good agreement with optical absorption, using $\varepsilon_{Cr(III)}$ equal to 18 L. mol$^{-1}$.cm$^{-1}$ ([16]).

The comparison of XANES and optical redox values (Fig. 4) indicates a good agreement between both methods. XANES may be used for redox determinations in Cr-doped glasses, although it is not possible to derive a calibration curve from physical mixtures of crystalline references [10, 21] [11]. Indeed, Cr(III) pre-edge feature is different in crystalline compounds and glasses (Table I), due to different Cr local environment.

The accuracy of redox values obtained by spectroscopic methods requires reference Cr(III) glasses of similar composition. However, without such ad hoc references, the uncertainty is lower when using XANES than for optical absorption spectroscopy, because the area variations of Cr(III) XANES pre-edge feature among glasses are smaller than $\varepsilon_{Cr(III)}$ variations derived from optical absorption spectroscopy . Moreover, at high Cr(VI) contents, the Cr(III) reference is less important. .

Using redox values derived from XANES spectroscopy allows to determine ε for Cr(VI) ($\varepsilon_{Cr(VI)}$) in partially oxidized glasses, for the 28 000cm$^{-1}$ band. A value of 4200±800 L.cm$^{-1}$.mol$^{-1}$ is obtained for SBN2 glass, in good agreement with the values published for silicate glasses, 4246 and 4218 L.cm$^{-1}$.mol$^{-1}$ [5], [18]. However, there is an important dispersion of the values, ranging between 5208 [22] and 7550 L.cm$^{-1}$.mol$^{-1}$[1]. The differences may be explained by the difficulty to prepare glass samples containing only Cr(VI). High $\varepsilon_{Cr(VI)}$ values imply the use of diluted samples low concentration or thin preparations, which increases the analytical uncertainty. $\varepsilon_{Cr(VI)}$ is also dependent on glass composition, such as in sodium borate glasses, in which it corresponds to a change of local Cr(VI) environment[23]. XANES appears to be an efficient method to measure Cr redox state in glasses.

## 4. Conclusion

The analysis of XANES pre-edge feature is a convenient method for the determination of the Cr(VI)/Cr$_{total}$ ratio. This method is in good agreement with optical absorption spectroscopy. Provided the use of appropriate references, especially at low Cr(III) content, XANES is a powerful method for the determination of Cr oxidation states in glasses.

## 5. Acknowledgments

The authors thank O. Proux for help with the XANES experiments, Marie-Hélène Chopinet for help with samples synthesis, the CAMPARIS service for microprobe analysis and the SARM of Nancy for chemical analysis.

Table I Molar compositions, molar extinction coefficients, XANES pre-edge peaks area and Cr(VI)/Cr$_{total}$ ratio of glasses determined by XANES and optical absorption.

| Glass samples | Composition, mol% | | | | | | $\varepsilon_{Cr(III)}$ (L. mol$^{-1}$.cm$^{-1}$) | Pre-edge area (%.eV) | Cr(VI)/Cr$_{total}$ | |
|---|---|---|---|---|---|---|---|---|---|---|
| | SiO$_2$ | Al$_2$O$_3$ | B$_2$O$_3$ | Na$_2$O | CaO | Cr$_2$O$_3$ | | | XANES | Opt. abs. |
| SN | 73.7 | 1.0 | 0 | 24.9 | 0 | 0.04 | 17.7±0.5 | 0.023 | 0 | 0 |
| SN2ox | 79.1 | 0 | 0 | 19.9 | 0 | 0.57 | 17.7 (a) | 0.087 | 27±3 | 30±6 |
| SCN | 73.2 | 1.0 | 0 | 14.3 | 10.8 | 0.04 | 18 (b) | 0.053 | 13±2 | 15±3 |
| SBN | 64.5 | 1.0 | 24.7 | 9.3 | 0 | 0.04 | 30.2±0.8 | 0.032 | 0 | 0 |
| SBNox | 66.0 | 1.1 | 24.1 | 8.4 | 0 | 0.04 | 30.2 (c) | 0.047 | 5±1 | 7±2 |
| SBN2 | 66.5 | 1.1 | 19.6 | 12.4 | 0 | 0.04 | 21.7±0.6 | 0.031 | 0 | 0 |
| BN | 0 | 0 | 66.7 | 32.0 | 0 | 0.36 | 18 (d) | 0.159 | 60±5 | 57±8 |

Impurities include 0.3% K$_2$O and 0.02% Fe$_2$O$_3$. Peak area is 0.012 for chromium alum and 0.247 for potassium bichromate (crystalline references).

(a) is taken from SN ; (b) from [1] ; (c) from SBN ; (d) from [16]

**Figure captions**

Fig. 1 Linear absorbance spectra of Cr(III) and Cr(VI) in SBN and SBNox glasses. The position of the absorption bands characteristic of the two Cr oxidation states is indicated.

Fig. 2 XANES spectra for glasses (bichromate reference in insert). Spectra are normalized to unity in the post-edge region.

Fig. 3 (left) XANES pre-edge features of glasses with a very low or no Cr(VI) content. (right) Pre-edge features of glasses with a high Cr(VI) content. Note the difference of relative intensity.

Fig. 4 Comparison of Cr(VI)/$Cr_{total}$ ratio determined by XANES and optical absorption methods. The origin point corresponds to Cr(III) bearing SN, SBN and SBN2 glasses.

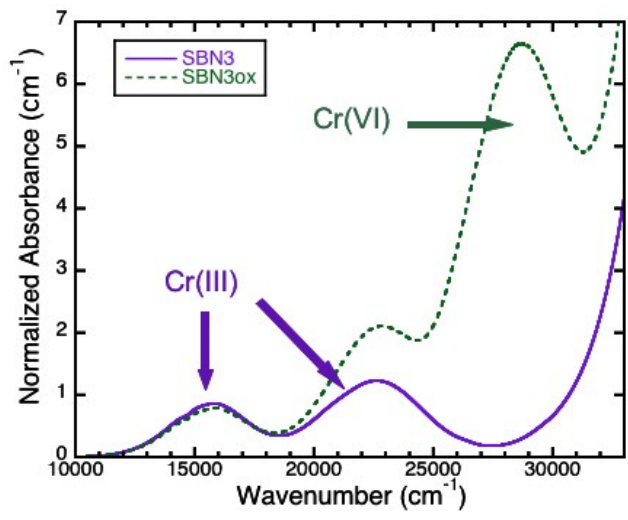

Figure 1

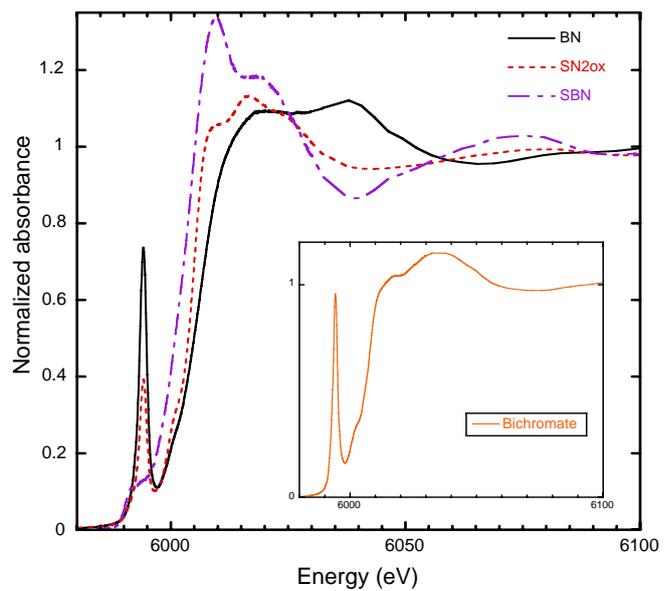

Figure 2

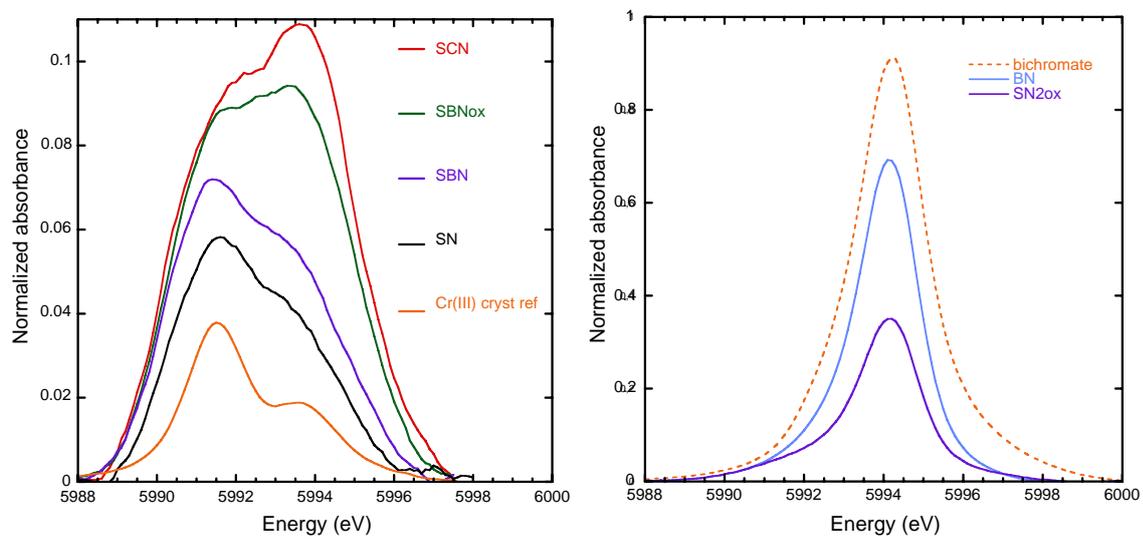

Figure 3

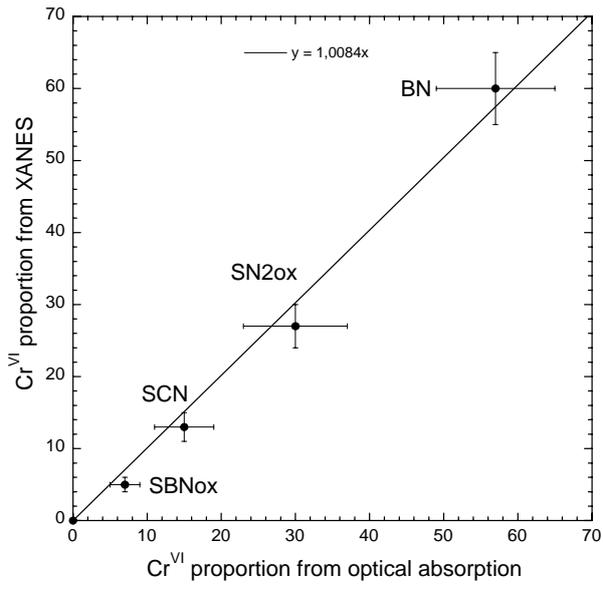

Figure 4